\documentstyle[12pt]{article}

\begin{document}


\newcommand{\ket}[1] {\mbox{$ \vert #1 > $}}
\newcommand{\bra}[1] {\mbox{$ <
#1 \vert $}}
\newcommand{\bk}[1] {\mbox{$ < #1 > $}}  \newcommand{\scal}[2]
{\mbox{$ < #1 \vert #2 > $}}
\newcommand{\expect}[3] {\mbox{$ \bra{#1} #2
\ket{#3} $}}  \newcommand{\ki}{\mbox{$ \ket{\psi_i} $}}
\newcommand{\bi}{\mbox{$ \bra{\psi_i} $}}  \newcommand{\p} \prime
\newcommand{\e}{\mbox{$ \epsilon $}}  \newcommand{\la} \lambda

\newcommand{\om}{\mbox{$ \omega $}}  \newcommand{\cc}{\mbox{$\cal C $}}
\newcommand{\w} {\hbox{ weak }} \newcommand{\al}{\mbox{$ \alpha $}}

\newcommand{\be}{\mbox{$ \beta $}}

\overfullrule=0pt \def\sqr#1#2{{\vcenter{\vbox{\hrule height.#2pt
          \hbox{\vrule width.#2pt height#1pt \kern#1pt
           \vrule width.#2pt}
           \hrule height.#2pt}}}}
\def\square{\mathchoice\sqr68\sqr68\sqr{4.2}6\sqr{3}6} \def\lrpartial{\mathrel
{\partial\kern-.75em\raise1.75ex\hbox{$\leftrightarrow$}}}

\begin{flushright}
LPTENS-95/2\\
February 1995
\end{flushright}
\vskip 1. truecm
\vskip 1. truecm

\centerline{\LARGE\bf{The Recoils of the Accelerated Detector}}
\vskip 2 truemm
\centerline{\LARGE\bf{and the Decohence of its Fluxes}}

\vskip 1. truecm
\vskip 1. truecm

\centerline{{\bf R. Parentani}\footnote{e-mail: parenta@physique.ens.fr}}
\vskip 5 truemm
\centerline{Laboratoire de Physique Th\'eorique de l'Ecole
Normale Sup\'erieure\footnote{Unit\'e propre de recherche du C.N.R.S.
associ\'ee \`a l'
ENS et \`a l'Universit\'e de Paris Sud.}}
\centerline{24 rue Lhomond
75.231 Paris CEDEX 05, France.}
\vskip 5 truemm

\vskip 1.5 truecm
\vskip 1.5 truecm
\vskip 1.5 truecm
\vskip 1.5 truecm

{\bf Abstract }
The recoils of an accelerated system caused by the transitions
characterizing thermal equilibrium are taken into account
through the quantum character of the
position of the system. The specific model is that of a two level
ion accelerated by
a constant electric field. The introduction of the wave function
of the center of mass position
clarifies the role of the classical trajectory in the
thermalization process. Then the effects of the recoils on the properties
of the emitted fluxes are analyzed. The decoherence of successive
emissions induced by the recoils
is manifest and simplifies the properties of the fluxes.
The fundamental reason for which one cannot neglect the recoils
upon considering accelerated systems is stressed and put in parallel
with the emergence of "transplanckian" frequencies in Hawking radiation.
\vfill \newpage
\vskip 1.5 truecm
\vskip 1.5 truecm

\section{Introduction}

It is now well known that an accelerated system perceives Minkowski
vacuum as a thermal bath\cite{Unr}. This can be understood from the fact that
the inner degrees of freedom of the system interact with the
local quanta of the radiation field. Indeed the accelerated
system interacts with the Rindler quanta and these quanta are
related to the usual Minkowski
ones by a non trivial Bogoliubov transformation\cite{Full}.

It also known that this Bogoliubov transformation is very similar in
character to the one that describes the scattering of modes in the time
dependent geometry of a collapsing star and  leads
to Hawking radiation\cite{Hawk}: a steady thermal flux.

In both cases, the kinematical agent
of the thermalization is treated classically.
Indeed in the accelerated situation, the uniformly accelerated trajectory
$z^2 - t^2 = 1/a^2$ is a priori given and unaffected by the recoils
of the system when it interchanges quanta during the transitions
characterizing thermal equilibrium.
Similarly, in the black hole situation, the background geometry is treated
classically and unaffected by the individual emissions of Hawking
quanta. Up to now the backreaction of Hawking radiation
onto the geometry has been achieved\cite{Bard}\cite{Massar2}\cite{PP}
only in the mean i.e. in the semi classical
approximation wherein only the mean value of the energy momentum
tensor enters in the Einstein equations treated classically.
This averaging procedure over individual emission acts appears to be
doubtful for the following reason\cite{THooft}\cite{Jacobson1}\cite{EMP}.
The frequencies involved in the
conversion from vacuum fluctuations to Hawking quanta rapidly exceed
the Planck mass. The reason for this
is precisely what leads to the thermalization
i.e. an exponentially varying Doppler factor relating incoming
scattered modes
to $out$-modes specified in the infinity future.
To understand the consequences of those transplanckian frequencies
for both the Hawking flux at latter times and  the
notion of a classical geometry is the major problem of black hole physics.

In the present article we shall investigate the much more modest problem
of the  effects of recoils of the accelerated system when one treats
its trajectory as a dynamical variable. That is, one introduces
an external field which causes the acceleration and one quantizes
the center of mass motion of the accelerated system.
The specific example treated here is that of a two level ion in a
constant electric field\cite{BPS}.
The consequencies of the recoils are the following.

The recoils do not modify the thermalization of the inner degrees of
freedom of the ion. The formulation in which the  center of mass
is quantized permits even to extend this thermalization to
more "quantum" situations wherein the accelerated system is described by
a completely delocalized wave function\cite{Parethese}.

Secondly the recoils do modify the properties of the flux emitted
by the accelerated system.
We first recall its properties in the absence
of
recoil\cite{UnrW}.
Once thermal equilibrium
is achieved, no mean flux is emitted\cite{grove}\cite{RSG}\cite{MPB}
 nor absorbed by the accelerated system,
in
perfect analogy with the usual inertial thermal equilibrium.
Nevertheless upon considering the total energy emitted one finds
a diverging energy\cite{Unr93}\cite{Aud}
 whose origin is that each internal transition
is accompanied by the emission of a Minkowski quantum. The reconciliation
between these seemingly contradictory results
lies in an analysis of the transients\cite{MaPa}
which shows how the interferences among all the Minkowski photons
destructively hide their presence in the intermediate equilibrium situation.
These interferences
are automatically washed out upon considering global quantities such as the
total energy emitted.

When the recoils of the accelerated
system are taken into account,
one finds a great simplification since the  Minkowski photons decohere
after a logarithmicly short proper time,
concomitant with the exponential Doppler shift mentioned above.
 Then the thermalized
accelerated system emits a steady
positive flux directly associated with the produced Minkowski quanta.
This steady emission of a positive flux
comes from the steady conversion of
potential energy in the electric field into radiation.

The paper is organized as follow.
In section 1, we briefly recall the properties of the
transition amplitudes and the fluxes in the no-recoil situation.
In section 2 we present the model of the two level ion. In section 3
we analyze the modifications of the transition amplitudes introduced by the
strict momentum conservation which accounts for the recoil.
Finally in section 4 we show how the finiteness of the rest mass of
the accelerated system leads, after a logarithmicly short time,
to the decoherence of the emitted fluxes.
We conclude the paper by some remarks mentioning other situations wherein
recoil effects play or might play an important role.

\section{The
Fluxes in the Absence of Recoil}

In this section, we present the relevant properties of the transition
amplitudes
which lead to the thermalization of the accelerated system\cite{Unr}. We
then analyze
the properties of the flux emitted by the accelerated
atom\cite{UnrW}\cite{grove}\cite{RSG}\cite{MPB}.
Since this material is now well understood and already
published\cite{Unr93}\cite{Aud}\cite{MaPa},
we shall
restrict ourselves to the presentation of the various properties
which will be compared to the ones obtained in the situation wherein the recoil
of the atom is taken into account.

We consider a two level atom maintained, for all times,
exactly on the uniformly accelerated trajectory
\begin{equation} t_a (\tau) =a^{-1}
\mbox{sinh} a \tau \ ,\  z_a (\tau) =a^{-1} \mbox{cosh} a \tau
\label{acctraj}
\end{equation}
where $a$ is the acceleration and $\tau$ is the proper time of the accelerated
atom. We work for simplicity in Minkowski
space time in $1+1$ dimensions.
The two levels of the atom are designated by $\ket{-}$ and $\ket{+}$ for the
ground state and the excited state respectively. The transitions from one state
to another are induced by the operators $A, A^{\dagger}$
\begin{eqnarray}
A \ket{-} = 0, \quad A \ket{+} = \ket{-} \nonumber\\
A^{\dagger} \ket{-} = \ket{+}, \quad A^{\dagger} \ket{+} = 0
\label{operA}
\end{eqnarray}

The atom is coupled to a massless scalar field $\phi(t,z)$. The
Klein-Gordon equation is $(\partial^2_t - \partial^2_z ) \phi =0$
 and the general solution is thus \begin{equation}
\phi(U,V) = \phi(U) + \phi(V) \label{phiuv}
\end{equation}
where $U,V$ are the light like coordinates given by $U=t-z, V=t+z$.
The general solution of the right moving part may be decomposed into
the orthonormal plane waves:
\begin{equation} \varphi_{\om} (U) = {e^{-i\om U} \over \sqrt{4 \pi \om}}
\label{modeU}
\end{equation}
where $\om$ is the Minkowski energy. The Hamiltonian of the right moving modes
is given by
\begin{equation}
H_M= \int_{-\infty}^{+\infty} \! dU T_{UU}
=  \int_0^{\infty} \! d\om {\om }
( a_{\om}^\dagger  a_{\om} )
\label{hamiltU}
\end{equation}
where $T_{UU} = (\partial_U \phi)^2$ is the normal ordered (with respect
to the Minkowski operators) flux and where
the Heisemberg field operator $\phi$ has been decomposed into
the Minkowski operators of destruction and
creation $a_{\om}, a_{\om}^\dagger$ as
\begin{equation}
\phi(U)= \int_0^{\infty} d\omega  \left( a_{\om}
\varphi_{\om} (U) + a_{\om}^\dagger \varphi_{\om}^* (U)\right)
\label{phiU}
\end{equation}
The operators $a_{\om}$ annihilates Minkowski vacuum $\ket{Mink}$, the ground
state of the Hamiltonian $H_M$.
In this article we shall restrict ourselves
to the $U$-part of the $\phi$ field only. (Note however
that  the coupling with the atom
introduces some mixing among the $U$ and the $V$-parts; but
this mixing does not
modify the properties of the flux nor the
conclusions of this article).

The coupling between the atom and the field is taken to be
\begin{equation}
\int\! dt dx\ \cal{H}_{\rm int} (t,x)  = \int\! d\tau H_{\rm int} (\tau)
=  g a \int\!
d\tau \left[ \left( A e^{-i \Delta m\tau} + A^{\dagger}
e^{i \Delta m\tau} \right)\phi(U_a(\tau)) \right]
 \label{Hint}
\end{equation}
where $U_a(\tau)=t_a(\tau) - z_a(\tau)= -e^{-a\tau}/a $
and where $H_{\rm int} (\tau)$ is
\begin{equation}
H_{\rm int} (\tau) = g  a \int_0^{\infty} \! {d\om
\over \sqrt{4 \pi \om}} \left[ \left(
A e^{-i \Delta m\tau} + A^{\dagger}
e^{i \Delta m\tau} \right) \left( a_{\om}{ e^{i{\om e^{-a\tau}}/a}
}
+ a_{\om}^\dagger {e^{-i\om e^{-a\tau}/a}
}
\right)\right]
\label{hamiltint}
\end{equation}
By the locality of the coupling, $\phi(U_a(\tau))$ evaluated only along
the classical trajectory enters into the interaction.

Then the transition amplitude (spontaneous excitation) from
$\ket{-}\ket{Mink}$ to the excited state $\ket{+}\ket{1_{\om}}$
(where $\ket{1_{\om}}= a^\dagger_{\om }\ket{Mink}$ is the one particle state
of energy-momentun $\om$) is
\begin{equation}
B(\om,\Delta m) = \bra{1_{\om}}\bra{+} e^{-i\int d\tau H_{\rm int}}
\ket{-}\ket{Mink}
\label{B}
\end{equation}
To first order in $g$ it is given by
\begin{eqnarray}
B(\om,\Delta m) &=& -ig a \int_{-\infty}^{+\infty} d\tau
e^{i \Delta m \tau} {e^{-i\om e^{-a\tau}/a} \over \sqrt{4 \pi \om}}
\nonumber\\
&=& ig
\Gamma(-i {\Delta m / a})
 {(\om/a)^{i{\Delta m / a}} \over \sqrt{4 \pi \om}}
e^{-\pi {\Delta m / 2a}}
\label{B1g}
\end{eqnarray}
where $\Gamma(x)$ is the Euler function.
This transition amplitude is closely related to the $\beta$ coefficient
of the Bogoliubov transformation\cite{Full}
 which relates the Minkowski operators
$a_{\om}$ to the Rindler operators associated with the eigenmodes of
$-i\partial_{\tau}$ (hence given by $\varphi_{Rindler}(\tau) = e^{-i\lambda
\tau}$). This relation indicates
that the accelerated atom absorbs (or emits) Rindler
quanta.

The transition amplitude (disintegration) from $\ket{+}\ket{Mink}$ to
the state $\ket{-}\ket{1_{\om}}$ is given by
$A(\om,\Delta m) = B(\om, -\Delta m)$. To first order in $g$ one finds
\begin{equation}
A(\om,\Delta m) = - B^*(\om,\Delta m) e^{\pi \Delta m /a}
\label{ratio}
\end{equation}
Thus for all $\om$ one has
\begin{equation}
\vert {B(\om,\Delta m) \over A(\om,\Delta m) }\vert^2 = e^{-2\pi \Delta m /a}
\label{ratio2}
\end{equation}
Since this ratio is independent of $\om$, the ratio of the probabilities of
transitions (excitation and disintegration) is also given by eq.
(\ref{ratio2}).
Hence at equilibrium, the ratio of the probabilities $P_-, P_+$
 to find the atom in the ground or excited state satisfy
\begin{equation}
{P_+ \over P_-}= \vert {B(\om,\Delta m) \over A(\om,\Delta m)} \vert^2 =
e^{-2\pi \Delta m /a}
\label{ratio3}
\end{equation}
This is the Unruh effect\cite{Unr}
: at equilibrium, the probabilities of occupation
are thermally distributed with temperature $a/2 \pi$.

In preparation for the "recoil case" and following ref. \cite{PaBr1},
we evaluate the
amplitude $B(\om, \Delta m)$ at the
saddle point approximation in order to localized around
which value of $\tau$ does the amplitude acquire its value.
The stationary phase of the integrand of eq. (\ref{B1g}) is at
\begin{equation}
\Delta m = - \om e^{-a \tau^*(\om)}
\label{sadd}
\end{equation}
hence at
\begin{equation}
\tau^*(\om) = {1 \over a } \ln {\om \over \Delta m} + i \pi /a
\label{saddtau}
\end{equation}
where the imaginary part is fixed by an analysis in the complex $\tau$ plane.
Similarly the stationary phase for the disintegration amplitude
$A(\om, \Delta m)$
gives $\Delta m =\om e^{-a \tilde \tau (\om)}$. Thus the saddle $
\tilde \tau$ is given
by the real part of eq. (\ref{saddtau}) (i.e. the saddle $ \tilde
\tau$ is the time
at which the exponentially varying Doppler factor $e^{-a \tau}$
brings $\om$ in resonance
with $\Delta m$). In both cases, the width
around the saddle is independent of $\om$ and
equal to $(\Delta m a)^{-1/2}$. This allows for the establishment of a rate
through successive resonances with different $\om$\cite{PaBr1}\cite{GO}.
Very important also is the fact that the difference ($=i \pi /a$)
between the saddles times for
$A$ and $B$ is imaginary and independent of $\om$.
Indeed, it leads to the fact
that $B(\om, \Delta m)$ and $A(\om, \Delta m)$, evaluated at
the saddle point ($s.p.$)
approximation,
satisfy automatically
\begin{eqnarray}
B_{s.p.}(\om, \Delta m)
 &=& -i g { (\om/ \Delta m)^{-i \Delta m /a} \over \sqrt{4 \pi \om}}
e^{- \pi \Delta m /a} \sqrt{{2 \pi a \over i \Delta m}} e^{ i \Delta m/a}
\nonumber\\
&=& - A^*_{s.p.} (\om, \Delta m) e^{- \pi \Delta m /a}
\label{basp}
\end{eqnarray}
Thus eq. (\ref{ratio2}) is satisfied and therefore, at equilibrium,
eq. (\ref{ratio3}) as well.
One should nevertheless add that the validity of the
saddle point approximation requires $\Delta m >> a $. Thus,
strictly speaking, this method is
valid only in the Boltzman regime where the density
of particles is low. We recall that the norm of $B(\om, \Delta m)$
given in eq. (\ref{B1g})
is proportional to the Bose Einstein distribution:
$(e^{2 \pi \Delta m /a} -1)^{-1}$.

We now analyze the properties of the mean flux emitted by the atom.
When the initial state is  $\ket{-}\ket{Mink}$, the mean flux emitted at
$V=\infty$ is given by
\begin{equation}
\langle T_{UU} \rangle = \bra{Mink} \bra{-}
e^{i\int d\tau H_{\rm int}} (\partial_U \phi)^2 e^{-i\int d\tau
H_{\rm int}}
\ket{-}\ket{Mink}
\label{tuu1}
\end{equation}
To order $g^2$, the state $e^{-i\int d\tau H_{\rm int}}\ket{-}\ket{Mink}$
is
\begin{eqnarray}
e^{-i\int d\tau H_{\rm int}}\ket{-}\ket{Mink} &=& \ket{-}\ket{Mink}
+\int_0^{\infty} d\om B(\om, \Delta m)
\ket{+} \ket{1_{\om}} \nonumber\\
&&+\int_0^{\infty} d\om \int_0^{\infty} d\om^\prime C(\om, \om^\prime, \Delta
m)
\ket{-} \ket{1_{\om} 1_{\om^\prime}}
\label{stateg2}
\end{eqnarray}
where \begin{equation}
C(\om, \om^\prime, \Delta m) =  \bra{1_{\om} 1_{\om^\prime}}
\bra{-} e^{-i\int d\tau H_{\rm int}} \ket{-}\ket{Mink}
\label{C}
\end{equation}
is the amplitude to emit two Minkowski photons.
Thus to order $g^2$, the mean flux is \begin{equation}
\langle T_{UU} \rangle = \langle T_{UU} \rangle_1 + \langle T_{UU} \rangle_2
\label{2terms}
\end{equation}
where \begin{equation}
\langle T_{UU} \rangle_1 = 2 \int_0^{\infty} d\om \int_0^{\infty} d\om^\prime
B(\om, \Delta m) B^*(\om^\prime, \Delta m) {\sqrt{\om \om^\prime}
\over 4\pi}
{e^{i(\om^\prime - \om)U}}
\label{T1}
\end{equation}
comes from the square of the linear term in $g$ of eq. (\ref{stateg2}), and
where
\begin{equation}
\langle T_{UU} \rangle_2 = - 2 \mbox{Re} \left[
\int_0^{\infty} d\om \int_0^{\infty} d\om^\prime
C(\om, \om^\prime, \Delta m)
{\sqrt{\om \om^\prime}
\over 4\pi}
{e^{-i(\om^\prime +\om)U}}
\right]
\label{T2}
\end{equation}
comes from the interference between the (unperturbed) first
 term of eq. (\ref{stateg2})
and the third term containing two particles.

Two important properties should be noted.
The first one  is that $\langle T_{UU} \rangle_2$ does not
contribute to the total energy emitted since $\int_{-\infty}^{+\infty}
dU e^{-i(\om^\prime +\om)U} = \delta (\om^\prime +\om)$ and
$\om, \om^\prime >0$. The total energy emitted is thus given by $
\langle T_{UU} \rangle_1$ only
\begin{equation}
 \langle H_M \rangle =
\int_{-\infty}^{+\infty}dU \langle T_{UU} \rangle_1 = \int_0^{\infty} d\om
\ \om \
\vert B(\om, \Delta m) \vert^2
\label{Energ}
\end{equation}
As remarked in \cite{Unr93}\cite{Pare}\cite{MaPa},
this energy diverges due to the U.V.
character of the amplitudes $B(\om)$ (We shall not be bothered by the I.R.
behavior since the contribution
for the total energy is always finite). Thus a cut off is needed in order to
obtain a finite energy. The simplest cut off consist on multiplying $B(\om)$
by
the regulator $e^{-\varepsilon \om}$.
By virtue of the locality of the resonance condition eq. (\ref{saddtau}),
the introduction of the regulator which damps the $\om > 1/\varepsilon$
mimics a switch off function around
\begin{equation}
\tau_{\varepsilon}= {1 \over a} \ln ({1 \over \varepsilon \Delta m})
\label{taufin}
\end{equation}
where $ \varepsilon$ is such that $\tau_{\varepsilon} >> 1/a$
(indeed the interaction should last many $1/a$ times in order for the atom
to properly thermalize\cite{MaPa}).

The second important property concerns the relative importance
of $\langle T_{UU} \rangle_1$ and $\langle T_{U
U} \rangle_2$ in the
stationary regime (when the Golden Rule is applicable i.e. when the
transition probability grows lineary with $\tau$)).
This intermediate behavior is manifest, for $|U|>\varepsilon$,
 when one compute explicitly
$\langle T_{UU} \rangle_1$ and $\langle T_{U
U} \rangle_2$ using the regulator $\varepsilon$.
Indeed one finds
\begin{eqnarray}
\langle T_{UU} \rangle_1 &=& + 2 \left({g \over 4 \pi }\right)^2
\vert \Gamma(i \Delta m /a)
\vert^4 \ \left[{1 \over U^2 + \varepsilon^2 }\right]
 \left[ \theta(U) + e^{-2\pi \Delta m /a} \theta(-U) \right]
\label{T11b}
\\
\langle T_{UU} \rangle_2 &=& - 2 \left({g \over 4 \pi }\right)^2
\vert \Gamma(i \Delta m /a)
\vert^4 \ \mbox{Re} \left[ {1 \over (U + i\varepsilon)^2 }
\right]
\label{T22b}
\end{eqnarray}
(The essential simplifying step in the computation of
$\langle T_{UU} \rangle_2$
is the replacement of $C(\om, \om^\prime, \Delta m)$ by
$A(\om^\prime, \Delta m)B(\om,\Delta m)$. This is legitimate when the Golden
Rule conditions are satisfied. See ref. \cite{MaPa} for the details.)

For positive $U$ (in the left quadrant where the atom isn't),
$\langle T_{UU} \rangle_1 + \langle T_{UU} \rangle_2 = 0$ as causality
requires\cite{UnrW}. For negative $U$, one finds (for $-U>\varepsilon$)
a steady absorption of Rindler energy\cite{grove} associated with the
steady increase of the probability to find the atom in the excited state
\begin{equation}
\langle T_{UU} \rangle_1 + \langle T_{UU} \rangle_2 = - 2\left( {g \over 4 \pi
}\right)^2
\vert \Gamma(i \Delta m /a)
\vert^4 (1 - e^{-2\pi \Delta m /a})
/U^2
\label{absorb}
\end{equation}

Thus one has two very different regimes. In the intermediate period
$\langle T_{UU} \rangle_2$ dominates the flux in accordance with the
Golden Rule description in the accelerated frame in which one sees the
absorption of a Rindler quantum. But when integrated, to find
the total Minkowski
emitted, see eq. (\ref{Energ}),
$\langle T_{UU} \rangle_2$ gives no contribution at all.
The reconciliation of the two
pictures arises from a detailed analysis of the transients\cite{MaPa}. For
instance one sees from eq. (\ref{T22b}) that
$\langle T_{UU} \rangle_2$ is positive
for $|U|<\varepsilon$ and one verifies that its integral over $U$
vanishes.

In a similar fashion one may study the equilibrium regime. One
includes the flux associated with the transition from
$\ket{-}$ to $\ket{+}$ and weights
the two contributions with the thermal occupation probabilities
given in eq. (\ref{ratio3}). One finds that in
thermal equilibrium, there is
no mean flux
as first pointed out by Grove\cite{grove}. Nevertheless,
in a global description,
when one compute the total energy radiated or the total number of
quanta radiated during the interacting period, one finds again a contribution
which seems to occur at a constant rate\cite{MaPa}.
Again, the reconciliation between the two descriptions
necessitates the analysis of
the transients when one switches off the interaction (i.e. for
$|U| \simeq \varepsilon$).

This concludes the analysis when the recoil of the atom is not taken
into account. We emphasize that all the emitted Minkowski quanta
interfere constructively in the $\langle T_{UU} \rangle_2$ term so as to
maintain a negative mean flux (eq. (\ref{absorb}))
in the Golden Rule regime. The
extremely well tuned phases which lead to this non vanishing character of
$\langle T_{UU} \rangle_2$ will be washed out after a finite proper time
when recoils will be taken
into account
through momentum conservation.

\section{The Two Level Ion in a Constant Electric Field}

In this section, we present the model introduced in ref.
\cite{BPS} (which is similar to the one used by Bell and
Leinaas\cite{BL})
which will allow us
to take automatically into account the recoils of the ion caused
by the transitions characterizing the accelerated thermal situation.
This model will also allow us
to prove that the thermalization of the inner degrees
of freedom of an accelerated system does not require
 a well defined
classical trajectory. Indeed, even when
one deals with delocalized waves
for the position of the ion,
the ratio eq. (\ref{ratio2}) is obtained\cite{Parethese} and
therefore the thermal equilibrium ratio eq. (\ref{ratio3}) is obtained
as well.

The model consist on two scalar charged fields ($\psi_M$ and $\psi_m$)
of slightly different
masses ($M$ and $m$)
 which will play the role of the former states of the atom: $
\ket{+}$ and $ \ket{-}$. The quanta of these fields are accelerated by an
external classical constant electric field $E$. One has
\begin{equation}
{E \over M} = a \simeq {E \over m}
\label{accE}
\end{equation}
because one imposes
\begin{equation}
\Delta m = M - m << M
\label{diffm}
\end{equation}
to have the mass gap well separated from the rest mass of the ion.

We work in the homogeneous gauge ($A_t=0$, $A_z= -Et$). In that gauge,
 the momentum $k$ is a conserved
quantity and the energy $p$ of a relativistic particle of mass $M$ is given by
the mass shell constraint $(p_{\mu}- A_{\mu})^2=M^2$ i.e.
\begin{equation}
p^2(M, k, t) = M^2 + (k + Et)^2
\label{KGE}
\end{equation}
The classical equations of motion are easily obtained from this equation
and are given in terms of the proper time $\tau$ by
\begin{eqnarray}
p(M,k, t) &=& M \mbox{cosh} a \tau \nonumber\\
t + k/E &=& (1/ a)  \mbox{sinh} a \tau \nonumber\\
z - z_0  &=& (1/ a) \mbox{cosh} a \tau
\label{eqmot}
\end{eqnarray}
Thus at fixed $k$, the time of the turning point (i.e. $dt/d\tau = 1$)
is fixed whereas its position is arbitrary.

{}From eq. (\ref{KGE}),
the Klein Gordon equation for a mode $\psi_{k,M}(t,z)=
e^{ikz} \chi_{k,M}(t)$ is
\begin{equation}
\left[ \partial_t^2 + M^2 + (k+Et)^2 \right] \chi_{k,M}(t) = 0
\label{KGE2}
\end{equation}
When $\Delta m \simeq a $ and when eq. (\ref{diffm}) is satisfied,
one has $M^2/E>>1$. Then the Schwinger pair production amplitude\cite{schw}
may be completely ignored (since the mean density of produced pairs
scales like $e^{-\pi M^2/E}$). Furthermore,
in this case,
the W.K.B. approximation for the modes  $\chi_{k,M}(t)$ is
valid for all $t$. Indeed, the corrections to this approximation
are smaller than $(M^2/E)^{-1}$.
\newpage
The modes
$\psi_{k,M}(t,z)$ can be thus correctly approximated by
\begin{equation}
\psi_{k,M}(t,z) = { e^{ikz}\over \sqrt{2 \pi}}{
e^{-i \int^t p(M, k, t^\prime ) dt^\prime} \over \sqrt{p(M, k, t)}}
\label{WKB}
\end{equation}
where $p(M, k, t^\prime )$ is the classical energy at fixed $k$
given in eq. (\ref{KGE}).

As emphasized in refs. \cite{BPS}\cite{BMPPS}, the wave packets of the form
\begin{equation}
\Psi_{k,M}(t,z)= \int dk^\prime {e^{-(k^\prime - k)^2/2 \sigma} \over
( \pi \sigma)^{1/4}} \psi_{k^\prime,M}(t,z)
\label{wp}
\end{equation}
do not spread if $\sigma \simeq E$.
In that case, the spread  (in $z$ at fixed $t$)
is of the order of ${E}^{-1/2}=(M a)^{-1/2}$ for all times and thus much
smaller than the acceleration length $1/a$ characterizing the classical
trajectory eq. (\ref{acctraj}). (This is readily seen by evaluating
the time dependence of
phase of $\Psi_{k,M}(t,z)$ at large $t$ when the W.K.B. approximation
becomes exact.) Since the stationary phase condition of the $\Psi_{k,M}(t,z)$
modes gives back the accelerated trajectory
and since
the wave packets do not spread, one has thus a center of mass position
quantized version
of the accelerated system, which furthermore tends uniformly
to the classical limit
when $M \to \infty, E \to \infty$ with $E/M= a$ fixed.

The interacting Hamiltonian which induces transitions between the quanta of
mass $M$ and $m$ by the emission or absorption of a massless neutral
quantum of the $\phi$ field is simply
\begin{equation}
H_{\psi \phi} = \tilde g M^2
\int dz \left[ \psi_M^\dagger (t,z)\psi_m (t,z) +
\psi_M(t,z) \psi_m^\dagger (t,z)
\right] \phi(t,z)
\label{inter2}
\end{equation}
where $\tilde g$ is dimensionless.
In momentum representation, by limiting ourselves to the right moving
modes of the $\phi$ field, one obtains
\begin{eqnarray}
H_{\psi \phi} = \tilde g M^2 \int_{-\infty}^{+\infty} dk \int_0^{\infty}
{ d\om \over \sqrt{4 \pi \om}}
\left[ b_{M,k-\om} \chi_{M,k-\om}(t) \ b^\dagger_{m,k}\chi^*_{m,k}(t)
+ \mbox{h.c.} \right] a_{\om} {e^{-i\om t} }
&&\nonumber\\
+\left[ b_{M,k+\om}\chi_{M,k+\om}(t) \ b^\dagger_{m,k}\chi^*_{m,k}(t)
+ \mbox{h.c.} \right] a_{\om}^\dagger {e^{+i\om t} }
&&\quad
\label{inter22}
\end{eqnarray}
where the operator $b_{M,k-\om}$ detroys
a quantum of mass $M$ and momentum
$k-\om$ and the operator $b^\dagger_{m,k}$ creates
a quantum of mass $m$ and momentum
$k$. Therefore the product $b_{M,k-\om}b^\dagger_{m,k}$ plays
the role of the operator $A$ (see eq. (\ref{operA}))
with, in addition, a strict conservation of
momentum. (We have not introduced anti-ion creation operators in the
Hamiltonian $H_{\psi \phi}$. This is a legitimate
truncation when $M^2/E >>1$.)

Contrary to what happens in
the original Unruh model, the interaction between the
radiation field $\phi$ and the two levels of the accelerated system is
no longer restricted a priori to a classical trajectory (see eqs. (\ref{Hint})
and (\ref{hamiltint})).
It is now in the behavior of the wave functions $\chi_{M,k}(t)$ that
the accelerated properties (and the thermalization after effects) are
encoded.

\newpage
\section{The Transition Amplitudes for the Two Level Ion}

As in the no-recoil model, we shall compute the transition amplitudes
as well as the properties of the fluxes caused by these transitions. The
analysis of the fluxes is presented in the next section.

We first prove that the behavior of the $\chi_{M}$ modes is
sufficient to obtain the Unruh effect and this without
having to localize the ion\cite{Parethese} i.e. without having to deal with
well localized wave packets.
Thus we compute the amplitude to jump from the state
$\ket{1_k}_m \ket{0}_M \ket{Mink}$ to the state
$\ket{0}_m \ket{1_{k^\prime}}_M \ket{1_{\om}}$. (Where $\ket{0}_M$ designates
the vacuum state for the $\psi_M$ field, and
where $\ket{1_k}_m = b_{m, k}^\dagger
\ket{0}_m$ is the one particle state of the $\psi_m$ field of
momentum $k$).
This amplitude is given by
\begin{eqnarray}
\cal{B}(k, k^\prime, \om) &=&  \bra{1_{\om}}_M \bra{1_{k^\prime}}_m \bra{0}
e^{-i \int dt H_{\psi \phi}} \ket{1_k}_m \ket{0}_M \ket{Mink}
\nonumber\\&=&  2\pi \delta (k - k^\prime - \om) \
\tilde B(\Delta m, k, \om)
\label{calB}
\end{eqnarray}
where momentum conservation occurs owing to the homogeneous character of the
electric field.
To first order in $\tilde g$,
$\tilde B(\Delta m, k, \om)$ is
\begin{eqnarray}
\tilde B(\Delta m, k, \om) &=& -i \tilde g M^2
 \int^{+\infty}_{-\infty}  dt \chi^*_{M, k - \om}(t) \chi_{m, k}(t)
{e^{i \om t} \over \sqrt{4 \pi \om}}
\nonumber\\&=&  -i \tilde g M^2
 \int^{+\infty}_{-\infty} dt  {e^{i\int^t dt^\prime
 \left[ p(M,k - \om, t^\prime) - p(m, k, t^\prime)
\right]} \over \sqrt{p(M,k - \om, t) p(m, k, t)}}
{e^{i\om t} \over \sqrt{4 \pi \om}}
\label{tildB2}
\end{eqnarray}
where we have used the W.K.B approximation eq. (\ref{WKB})
for the
$\chi$ modes.

To grasp the content of this amplitude, is it useful to first develop
the integrand in powers of $\om$ and $\Delta m$.
To first order in $\Delta m$ and $\om$, the phase $\varphi(\tau)$
of the integrand is
\begin{eqnarray}
\varphi(\tau) &=& \int^t \left[ p(M,k - \om, t^\prime) - p(m, k, t^\prime)
\right] + \om t
 \nonumber\\
&=&\Delta m \int^t dt^\prime
\partial_M p(M,k , t^\prime) - \om \int^t
dt^\prime
\partial_k p(M,k , t^\prime) + \om t  \nonumber\\ &=&
\Delta m \Delta \tau(t) - \om ( \Delta z_a(t) - t)
+ C
\nonumber\\ &=&
\Delta m \Delta \tau(t) - {\om \over a } e^{-a \Delta \tau(t)} - {\om k \over
E}
+ C
\label{phasa}
\end{eqnarray}
where $C$ is a constant and where $\Delta \tau(t)$ and $\Delta z_a(t)$ are
the classical relations, eqs. (\ref{eqmot}),
 between the lapses of proper time and of $z$ and $t
+ k/E$
evaluated along any uniformly accelerated trajectory.
Eq. (\ref{phasa}) may be checked explicitly by computing the integrals
using eq. (\ref{KGE}).
It is perhaps
more instructive to realize that those relations are nothing but the
Hamilton-Jacobi relations between the classical action $S_{cl.}$
and proper time or
momentum: $\partial_M S_{cl.} = - \Delta \tau$ and $ \partial_k S_{cl.} = -
\Delta z$. Hence, whatever is the nature of the external field which brings
the system into constant acceleration, the first two terms of eq. (\ref{phasa})
will always be found.

In addition, in that approximation,
one can neglect
the dependence in $\om$ and $\Delta m$
in the denominator of eq. (\ref{tildB2}).
Hence the measure
is
$dt/p(M,k, t) = d\tau /M$. Thus to first order in $\om$ and $\Delta m$,
one has \begin{eqnarray}
\tilde B(\Delta m, k, \om) &=& {-i \tilde g M}  \int^{+\infty}_{-\infty} d\tau
e^{i\Delta m \tau} {e^{-i \om e^{-a\tau}/a} \over \sqrt{4 \pi \om}}
e^{-i ( \om k/E - C)}
\nonumber\\
 &=& \left[ {\tilde g M \over ga} \right]
B(\Delta m, \om) e^{-i (\om k/E - C)}
\label{tildB3}
\end{eqnarray}
where $B(\Delta m, \om)$ is the amplitude in the no-recoil original
situation given in eq. (\ref{B1g}).
Very important is the fact that the initial momentum $k$ introduces only
a phase
in the amplitude $\tilde B(\Delta m, k, \om)$.
Thus any superposition of modes $\psi_{m,k} $ will give rise to the same
probability to emit of photon of energy $\om$.  And
the norm of the ratio
of $\tilde B(\Delta m, k, \om)$ over
$\tilde A (\Delta m, k, \om)$
satisfies eq. (\ref{ratio}). Therefore, in that approximation,
the two level ion
thermalizes exactly as in the no-recoil case.

Having understood the role of the various factors of the amplitude
$\tilde B(\Delta m, k, \om)$, one may now
refine the derivation and compute $\tilde B(\Delta m, k, \om)$
at the saddle point approximation
without developing a priori the integrand in powers of $\om$ and $\Delta m$.
The upshot of this calculation is that eq. (\ref{tildB3}) is correct up
to an additional phase factor and correction terms of the form
$(\Delta m/ M)^n\ O(1) $. In order for this to be true,
it is necessary to verify that terms in $(\Delta m/ M)^n$ are not
multiplied by factors scaling like $e^{a \tau}$, which one might have
feared owing to the scaling given in eq. (\ref{sadd})
of the resonant energy $\om$ with $\tau$.
This is an important condition because
the proper time necessary to get a rate formula, which gives rise to the
thermal distribution,
requires $\tau >> 1/a$.
A scaling factor $e^{a \tau}$ would then ruin
the condition that relies on $\Delta m/ M << 1$.

The stationary point $t^*$ of the integrand of eq. (\ref{tildB2}) is at
\begin{equation}
p(M,k - \om, t^*) - p(m, k, t^*) + \om = 0
\label{sp2}
\end{equation}
i.e. conservation of the Minkowski energy (contrary to the condition
eq. (\ref{sadd})
which was the resonance condition in the accelerated frame, i.e.
conservation of Rindler energy).
Taking the
square of eq. (\ref{sp2}) and using eq. (\ref{KGE}) one gets
\begin{equation}
{M^2 - m^2 \over 2 } = \om \left[ ( k - \om +Et^*) - p(M,k - \om, t^*)\right]
\label{sp2b}
\end{equation}
Introducing once more the proper time
eq. (\ref{eqmot}) one finds
\begin{equation}
\Delta m (1 - \Delta m / 2M) = - \om e^{-a \tau^*}
\label{sp2c}
\end{equation}
Thus, in our situation with
$\Delta m /M <<1$, the stationarity condition gives back
the resonance time eq. (\ref{saddtau}) (as well as its imaginary part)
 of the no-recoil model. A similar analysis of the inverse transition
$\tilde A(\Delta m, k, \om)$
(i.e. where the initial state $M$ has momentum $k$) leads back to
the saddle point $\tilde \tau$ given by eq. (\ref{sp2c})
with a $+$ sign on the r.h.s.

When evaluated at the saddle time $t^*$, the total phase
$\tilde \varphi(t^*)$ of $\tilde B$, eq. (\ref{tildB2}), is
\begin{eqnarray}
\tilde \varphi(t^*) &=& \om t^* + \int_0^{t^*+(k-\om)/E} dt^\prime
p(M, 0, t^\prime) - \int_0^{t^*+k/E} dt^\prime
p(m, 0, t^\prime)
\nonumber \\
&=& \om t^* + \int_0^{t^*+k/E} dt^\prime
\left[ p(M ,0, t^\prime) - p(m,0,  t^\prime)
\right] - \int^{ t^*+k/E}_{t^*+(k-\om)/E} dt^\prime
p(M, 0, t^\prime)
\nonumber \\
&=& \varphi_1(t^*) + \varphi_2(t^*) + \varphi_3(t^*)
\label{phase}
\end{eqnarray}
(We have fixed the constant $C$ of eq. (\ref{phasa}) by this
choice of the lower
bounds for the two
integrals. The same choice of the phase at different
$k$ is available when one construct a wave packet as in eq. (\ref{wp})). One
develops the second term in powers of $\Delta m /M$ and gets
\begin{equation}
\varphi_2(t^*) = {M^2 \over E} \left[ {\Delta m
\over M} ( 1 - {\Delta m
\over 2 M}) (a \tau^*) + ({\Delta m
\over M})^2 O(1)
\right]
\label{phase2}
\end{equation}
where all the higher powers of $\Delta m
/M$ are multiplied by $O(1)$. Thus $\varphi_2(t^*) = \Delta m \tau^*$.
The third term is developed in powers of $\om $ and reads
\begin{eqnarray}
\varphi_3(t^*) &=& - { \om \over E} \left[ p(M, k, t^*) - {\om \over
2} {k + Et \over p(M, k, t^*)} + {\om^2 \over
p(M, k, t^*)} O(1) \right] \nonumber \\
&=& - \om \Delta z(t^*) + { \om ^2 \over 2E} \mbox{tanh} (a\tau^*)
\left( 1 +  {\Delta m   \over M} O(1) \right)
\label{phase3}
\end{eqnarray}
where we have used eq. (\ref{eqmot}) to replace $p/E$ by $\Delta z$.
We have also used eq. (\ref{sp2c}) to replace $\om /p(M, k, t^*)$
by $(\Delta m/M) O(1)$.
In addition, one verifies that
all the higher powers of $\om$ appear in the form $(\om/p)^n$ only, hence,
by the same replacement, are expressible as
$(\Delta m/M)^n \ O(1)$ and are therefore negligible.
The fact that all powers of $\om$ appear in the ration $\om/p$ is not an
accident. It means that in the boosted frame at rest at $\tau = \tau^*$,
the recoil effects are controlled by $\Delta m/M$. We believe that this
property will be found whatever is the accelerating external field.

Only the second term cannot be neglected since the resonant $\om$
($= \Delta m e^{a \tau^*}
$) is bigger than $E^{1/2}$ after a proper time $\tau_{recoil}$
equal to
\begin{equation}
\tau_{recoil} = { 1 \over 2a} \ln { M \over \Delta m} + { 1 \over 2a} \ln {
a \over \Delta m}
\label{taur}
\end{equation}
This is related to the fact that it is kinematically inherent
for uniformly accelerated systems that the energy $\om$ exchanged
during transitions exceeds the rest mass of the system in a
logarithmicly short proper time
given by $a\tau=\ln  M / \Delta m$. In the collapsing black
hole situation, for exactly the same kinematical reasons
(i.e. the formation of an horizon giving rise to an exponentially varying
Doppler factor),
 one finds that the resonant
frequencies $\om$ which give rise to the Hawking quanta exceed the
Planck mass\cite{THooft}\cite{Jacobson1}\cite{EMP}\cite{MaPa}
 after a logarithmicly short time (measured at spatial
 infinity).

The quadratic
spread $\langle \Delta t \rangle^{2}$
around the saddle $t^*$, which is given by the inverse of
the second derivative of the phase, is (for both $\tilde B$ and $\tilde A$)
\begin{eqnarray}
\langle \Delta t \rangle^{-2}
&=& E \left[ { k - \om +Et^* \over p(M,k - \om, t^* )} -
 { k +Et^* \over p(m, k,t^*) } \right]
\nonumber \\
&=& E \left[{ (M^2 - m^2)/2 \over p(M,k - \om, t^*)
p(m, k,t^*) } \right]
 \nonumber \\
&=& \langle \Delta \tau  \rangle^{-2}
 ({dt \over d\tau})^{-2} \left[ 1 + {\Delta m
\over M} O(1) \right]
\label{spread}
\end{eqnarray}
where $\langle \Delta \tau \rangle $ is the spread of proper time
in the no-recoil situation.
Similarly the denominator of eq. (\ref{tildB2}) is
\begin{equation}
{dt \over \sqrt{ p(M,k - \om, t^*)
p(m, k,t^* )}} = d \tau \left[ 1 + {\Delta m
\over M} O(1) \right]
\label{det}
\end{equation}
where powers of $\om /p(M, k, t^*)$ are powers of $\Delta m /M $
hence negligible.

By collecting the various terms one obtains that $\tilde B (\Delta m, k, \om)
$ is given, at the saddle point approximation by
\begin{equation}
\tilde B_{s.p.} (\Delta m, k, \om) = \left[ {\tilde g M \over ga} \right]
B_{s.p.}(\Delta m, \om)
e^{-i \left[2 \om k -  { \om ^2 } \mbox{tanh} a\tau^* \right]/2E}
\label{calB22}
\end{equation}
where $B_{s.p.}(\Delta m, \om)$ is
given in eq. (\ref{basp}).
By a similar analysis, one easily shows that
\begin{equation}
\tilde A_{s.p.} (\Delta m, k, \om) = \left[ {\tilde g M \over ga} \right]
A_{s.p.}(\Delta m, \om)
e^{-i \left[2 \om k -  { \om ^2 } \mbox{tanh} a\tilde \tau \right]/2E}
\label{calA22}
\end{equation}
where $\tilde \tau = \mbox{Re} [\tau^*]$.
Since the modification for both amplitudes is a pure phase,
eq. (\ref{ratio}) is still satisfied. Hence at equilibrium one will
find the thermal population eq. (\ref{ratio3}) as well.
Strictly speaking we have shown the thermal equilibrium only when
$\Delta m >> a$ which is a necessary condition to have
 a good saddle point approximation.
But the fact that the differences of the integrands
of the amplitudes, between the present case
and the no-recoil case, scale like $\Delta m /M$
and not like $\Delta m /a$ indicates
that when $\Delta m \simeq a $ the square of the amplitude $\tilde B$
will be proportional to the full
Planck distribution as well\footnote{This has been verified explicitly
by S. Massar who used integral representations (see ref. \cite{BMPPS})
of the exact
solutions of eq. (\ref{KGE2}). He found that the norm of
$\tilde B$ is, as in the no-recoil case,
proportional to the Planck distribution.
Hence, the saddle point restriction $\Delta m >>a$
can be waived. Furthermore, since he worked with the exact solutions
of eq. (\ref{KGE2})
rather that W.K.B. approximations, it proves that, at least for the
ion in a constant electric field, the
thermalization of the inner degrees of freedom
does not even require the semi-classical limit: $M^2>>E$, see ref. \cite{BPS}.}
(see discussion after eq. (\ref{basp})).

\section{The Flux Emitted by the Two Level Ion}

We now compare the flux emitted by the two level ion, with the former
no-recoil flux given in eqs. (\ref{tuu1}$\to$\ref{T2}).
To this end we have to use well localized
wave packets.
To be the closest to the no-recoil case, we use the "minimal"
wave packet\cite{BPS}\cite{BMPPS}
given in eq. (\ref{wp}) with $\sigma = E$ to describe the initial
state of the ion of mass $m$. For simplicity
we shall center its mean momentum at $k=0$ and works with the phase
specified in eq. (\ref{phase}). Then the position of the turning point of the
mean trajectory encoded in the
initial wave function is at $t=0, z_0 =0$, see eq. (\ref{eqmot}).

To order $\tilde g^2$, the
flux is, in total analogy with eqs. (\ref{2terms}), (\ref{T1}) and
(\ref{T2}), given by
two terms
\begin{equation}
\langle \tilde T_{UU} \rangle = \langle \tilde T_{UU} \rangle_1 +
\langle \tilde T_{UU} \rangle_2
\label{T1+2}
\end{equation}
With the initial state of the ion $m$ given
as specified just above, $\langle \tilde T_{UU} \rangle_1$ is
\begin{eqnarray}
\langle \tilde T_{UU} \rangle_1 = \int dk_1 \int dk_2 {e^{-(k_1)^2/2 E}
\over
( \pi E)^{1/4}}
{e^{-(k_2)^2/2 E}
\over
( \pi E)^{1/4}}
\int_0^{\infty} d\om \int_0^{\infty} d\om^\prime
\delta( k_1 - \om - (k_2 - \om^\prime))&&
\nonumber\\
  \quad \quad  \quad \quad \quad  \left[2
\tilde B (\Delta m, k_1, \om) \tilde B^*(\Delta m, k_2, \om^\prime)
{\sqrt{\om \om^\prime}
\over 4\pi}
{e^{i(\om^\prime - \om)U}}
\right]
&&
\label{T13}
\end{eqnarray}
where the $\delta$ of Dirac comes from momentum conservation of
the exchanged ion
of mass $M$.
Similarly $\langle \tilde T_{UU} \rangle_2$ is
\begin{eqnarray}
\langle \tilde T_{UU} \rangle_2 =
\int dk_1 \int dk_2 {e^{-(k_1)^2/2 E}
\over
( \pi E)^{1/4}}
{e^{-(k_2)^2/2 E}
\over
( \pi E)^{1/4}}
\int_0^{\infty} d\om \int_0^{\infty} d\om^\prime
\delta( k_1 - \om - \om^\prime - (k_2))
&&\nonumber\\
  \quad \quad  \quad \quad \quad  (-2) \mbox{Re} \left[
\tilde C(k_1, \om, \om^\prime, \Delta m)
{\sqrt{\om \om^\prime}
\over 4\pi}
{e^{-i(\om^\prime +\om)U}}
\right]
&&\label{T23}
\end{eqnarray}
where $\tilde C(k_1, \om, \om^\prime, \Delta m)$ is the amplitude
to emit two photons of frequencies $ \om$ and $\om^\prime$
starting from the ion in the state $\ket{1_{k_1}}_m$. Notice that the
argument of the $\delta$ of Dirac is not the same
as in
eq. (\ref{T13}). This is due to the fact that it comes now from the overlap
between the twice scattered
ion $m$ of momentum $k_1 - \om - \om^\prime$ with the
unperturbed one of momentum $k_2$. This difference of arguments
will have a determinative
importance in the sequel.

Performing the $k_2$ integration one obtains
\begin{eqnarray}
\langle \tilde T_{UU} \rangle_1 &=&
2 \int_0^{\infty} d\om \int_0^{\infty} d\om^\prime
\int dk_1 {e^{-(k_1)^2/2 E}
\over
( \pi E)^{1/4}}
{e^{-( k_1 - \om + \om^\prime)^2/2 E}\over
( \pi E)^{1/4}}\nonumber\\
&&\quad \quad \quad \left[
\tilde B (\Delta m, k_1, \om) \tilde B^*(\Delta m, k_1 - \om + \om^\prime,
\om^\prime)
{\sqrt{\om \om^\prime}
\over 4\pi}
{e^{i(\om^\prime - \om)U}}
\right]
\label{T14}
\end{eqnarray}
and
\begin{eqnarray}
\langle \tilde T_{UU} \rangle_2 &=&
- 2 \int_0^{\infty} d\om \int_0^{\infty} d\om^\prime
\int dk_1 {e^{-(k_1)^2/2 E}
\over
( \pi E)^{1/4}}
{e^{-( k_1 - \om - \om^\prime)^2/2 E}\over
( \pi E)^{1/4}}
\nonumber\\
&& \quad \quad \quad    \mbox{Re} \left[
\tilde B (\Delta m, k_1, \om)
\tilde A (\Delta m, k_1 - \om ,\om^\prime)
{\sqrt{\om \om^\prime}
\over 4\pi}
{e^{-i(\om^\prime +\om)U}}
\right]
\label{T24}
\end{eqnarray}
(where we have replaced $\tilde C(k_1, \om, \om^\prime, \Delta m)$ by
$\tilde B (\Delta m, k_1, \om)
\tilde A (\Delta m, k_1 - \om ,\om^\prime)
$ c.f. the remark after eq. (\ref{T22b})).
Again one sees the difference of the recoil effects in the
different arguments of the second gaussian factor in eqs. (\ref{T14}) and
(\ref{T24}).

Analyzing the content of these fluxes, we note, first of all,
that, as in the no recoil case, $\langle \tilde T_{UU} \rangle_2$
does not contribute to the total energy emitted (see eq. (\ref{Energ})).
Furthermore
since the total energy is a function of the norm of
$\tilde B (\Delta m, k_1, \om)$ only (this is because only
$\om=\om^\prime$ contributes in eq. (\ref{Energ})) and since the
amplitudes $\tilde B (\Delta m, k_1, \om)$ differ from $B (\Delta m,\om)$
by a phase only, see eq. (\ref{calB22}), the
energy emitted is identical
to the one in the no-recoil case. The effect of the recoil is therefore
to modify the repartition of the energy density at most.

To verify that the recoil does modify
the repartition of the  energy density, we shall
use the saddle point approximation expressions, eqs.
(\ref{calB22}, \ref{calA22}),
for $\tilde B (\Delta m, k_1, \om)$
and for $\tilde A(\Delta m, k_1 - \om ,\om^\prime)$.
This leads to a tedious but straightforward computation.

We first simplify the $k$ dependence of the additional phase factor
${i (\om^2/2E) \mbox{tanh} a \tau^* }$ of eqs.
(\ref{calB22}, \ref{calA22}) where $\tau^* $ is a function $k$ through
eq. (\ref{eqmot}). We drop this $k$ dependence and
replace this phase by ${i \om^2/2E}$. This is legitimate since for
large $\tau$ the tanh$a\tau^*$ is exponentially
close to $1$, and for small $\tau^*$, the
resonant frequency $\om$ is of the order of $\Delta m$, hence this
phase is negligible anyway.

With this simplification, we can perform the $k_1$ integration
since it is now gaussian, and we obtain
\begin{eqnarray}
\langle \tilde T_{UU} \rangle_1 &=&
2 \int_0^{\infty} d\om \int_0^{\infty} d\om^\prime
e^{-(\om - \om^\prime)^2/2 E}
\nonumber\\
&& \quad \quad \quad \left[
 B_{s.p}(\Delta m,\om)B^*_{s.p.}(\Delta m,\om^\prime)
{\sqrt{\om \om^\prime}
\over 4\pi}
{e^{i(\om^\prime - \om)U}}e^{-(\om^\prime+\om)\varepsilon}
\right]
\label{T15}
\end{eqnarray}
and
\begin{eqnarray}
\langle \tilde T_{UU} \rangle_2 &=&
- 2 \int_0^{\infty} d\om \int_0^{\infty} d\om^\prime
e^{-(\om + \om^\prime)^2/2 E}
\nonumber\\
&& \quad \quad \mbox{Re} \left[
B_{s.p}(\Delta m,\om)A_{s.p.}(\Delta m,\om^\prime){\sqrt{\om \om^\prime}
\over 4\pi}
{e^{-i(\om^\prime +\om)U}}e^{-(\om^\prime+\om)\varepsilon}
\right]
\label{T25}
\end{eqnarray}
where the quantities which appear in the brackets are computed in
the no-recoil case (see eqs. (\ref{T1}, \ref{T2})). The regulator
$\varepsilon$ is given in eq. (\ref{taufin}) and we have set
$[\tilde g M/ ga]= 1 $ for simplicity.
 One sees that the effect of
the combined gaussian weights of the wave packet describing the initial state
of the ion  and the $k$ dependence
of the phase factors of the amplitudes $\tilde B$ and $\tilde A$
is to erase the $\om^2$ phase factors of eqs.
(\ref{calB22}, \ref{calA22}) and to introduce a gaussian factor into the
integrand of the no-recoil case. (One trivialy verifies that in the limit
$M \to \infty, E \to \infty$ with $E/M=a$ fixed, eqs. (\ref{T15}, \ref{T25})
give back identically eqs. (\ref{T11b}, \ref{T22b}) with $B$
and $A$ evaluated at the saddle point approximation.)
These gaussian factors
are the sole effect of the recoil on the energy repartition. They have the
following consequences.

To understand the $U$ dependence of
$\langle \tilde T_{UU} \rangle_1$ and $\langle \tilde T_{UU} \rangle_2$,
 let us first make clear the hierarchy of
characteristic proper times
and, by virtue of the resonance condition eq. (\ref{sp2c}),
the hierarchy of the
energies $\om$ exchanged. One has
\begin{equation}
\tau_0 < \tau_{recoil} + \tau_0 < \tau_{\varepsilon}  + \tau_0
\label{taus}
\end{equation}

$\tau_0$ is the time at which one prepares the initial wave packet,
it marks therefore the beginning of the interaction and
effectively defines the "laboratory Minkowski frame"
 in which the frequencies
$\om$ have an absolute meaning.
This is possible because the construction of the wave packets breaks
the Rindler
invariance present in the no-recoil case.
Indeed one can explicitly check that the wave packets given in eq. (\ref{wp})
are not invariant under boost. Using the wave packet given in eq.
(\ref{wp}) with $k=0$ and with the phase given in eq. (\ref{phase})
means that the construction is made at (around) $t=0$, hence in this case
$\tau_0 = 0$.

$\tau_{recoil}$ designates the lapse of proper time which starts at
$\tau_0$ and  indicates the moment at which the resonant
frequency $\om$ (measured in the Minkowski frame where $\tau_0 =0$)
is equal to $(Ma)^{1/2}$ (see eq. (\ref{taus}),
i.e. when the energy exchanged equals the spread in
$k$ of the wave packets of eq. ({\ref{wp}) with $\sigma = E$).
$\tau_{recoil}$ characterizes, as we shall see, the time from which recoil
effects dominate the physics. Furthemore, when $\Delta m \simeq a $, it is half
the time at which $\om$ equals the rest
mass of the ion. Had we used therefore another external field to put the system
into acceleration, we would have found a similar proper time
associated with the width of the wave packet.

$\tau_{\varepsilon}$ designates the end of the interaction. We recall
that it is
necessary to limit the frequency $\om$ in the U.V.

One has two different regimes. For $- U= e^{-a \tau}/a >> E^{-1/2}$ (i.e.
$\tau$ smaller than
$\tau_{recoil}$)
the resonant energy $\om$ is much smaller that $(aM)^{1/2}$, hence the
gaussian weights in eqs. (\ref{T15}, \ref{T25}) play no role and
the energy repartition $\langle \tilde T_{UU} \rangle_1 +
\langle \tilde T_{UU} \rangle_2$
is still the same as in the no-recoil case hence given by eq. (\ref{absorb}).
This can be immediately verified by computing the $\om$, $\om^\prime$ integrals
at the saddle point approximation. One finds indeed
that the saddle frequency $\om^*$
is at $ \om^* = -\Delta m/aU$.

Instead, for $- U< E^{-1/2}$ (i.e. $\tau > \tau_{recoil} $), the effects
of the gaussian factors on
$\langle \tilde T_{UU} \rangle_1$ and $
\langle \tilde T_{UU} \rangle_2$ are totally differently.
$\langle \tilde T_{UU} \rangle_2$ does no longer  scale like
$1/U^2$ (which expresses a constant rate in Rindler time).
Indeed, using eq. (\ref{basp}), eq. (\ref{T25}) reads
\begin{eqnarray}
\langle \tilde T_{UU} \rangle_2 &=&
- 2 \int_0^{\infty} d\om \int_0^{\infty} d\om^\prime
e^{-(\om + \om^\prime)^2/2 E}
\nonumber\\
&&
\quad \quad \left(- g^2 \left({ a\over \Delta m } \right)e^{-\pi \Delta m/a}
\right) \mbox{Re} \left[
\left({ \om \over \om^\prime }\right)^{-i \Delta m/a}
e^{-i(\om^\prime +\om)U} e^{-(\om^\prime +\om)\varepsilon}
\right]
\nonumber\\
&=&
+ 2 g^2 \left({  \Delta m \over a } \right)
e^{-2\pi \Delta m/a}
\int_0^{\infty} d\om_1 \ \om_1 \ \mbox{cos}(\om_1 U)
e^{- \om_1^2/2 E} e^{- \om_1 \varepsilon}
\label{t2final}
\end{eqnarray}
where we have defined $\om_1 = \om + \om^\prime$ and integrated over the
angle arctan$(\om/\om^\prime)$.
The behavior of $\langle \tilde T_{UU} \rangle_2$ at small $U$ is now
governed by $E$ and no longer by $\varepsilon$. Indeed the development
of the integral at small $U$ gives
$E(1 - EU^2 + O(EU^2)^2)$ instead of $-$Re$[1/(U+i \varepsilon)^2]$
obtained in
the no
recoil case in eq. (\ref{T22b}).

On the contrary, the behavior of $\langle \tilde T_{UU} \rangle_1$
has to be still like $1/U^2$ since the total energy emitted
is unaffected by the recoil and scales like $1/\varepsilon$.
This is confirmed by the saddle point approximation of the $\om$,
$\om^\prime$ integrations.

Thus for $-U<E^{-1/2}$, the negative
contribution of $\langle \tilde T_{UU} \rangle_2$ becomes completely
negligible as compared to the unaffected $\langle \tilde T_{UU} \rangle_1$
contribution. Therefore one finds that the flux emitted by the accelerated
system is now positive and given by the contribution of the first term in
the Born series (linear in $g$, i.e. the
$B$ term in eq. (\ref{stateg2})).

Similarly, in the equilibrium
situation, for $\tau >\tau_{recoil}$,
the flux is given
by the sum of the positive contributions of the excitation and disexcitation
processes.
One has then that all transitions characterizing thermal equilibrium
lead to a positive flux in situ contrariwise of what happened in the no
recoil case owing to the interferences among all emitted quanta.
Therefore, the
decoherence greatly simplifies the resulting flux
which is now much closer to what one might have
naively expected (i.e. one should no longer be bothered by the formerly
interfering $\langle T_{UU} \rangle_2$ terms).

Furthermore, there is a strict relation between this decoherence and the
conservation
of momentum, eq. (\ref{calB}), and energy, eq. (\ref{sp2}).
Indeed, the two level ion constantly loses energy and momentum
in accordance with these conservation laws.
To understand how momentum conservation
modifies the mean trajectory of the ion, compare the orbit where
there is no recoil ($g \to 0$)
with the orbit of the emitting two level ion.
There is, in the second case, a succession of hyperbolae such that their
turning points (see eqs. (\ref{eqmot})) drift in $t$ and $z$ corresponding
to laters times and greater $z$. The total change in the time of
the turning point is $ E \Delta t_{t.p.}= \sum_i \om_i$
(i.e. the total momentum lost
is the sum of the successive losses due to each transition).
One also verifies that the total change in position is $\sum_i \om_i/E$.
These successive changes of hyperbolae\footnote{
P. Grove and D. Raine emphasize that one should consider
a "Rindler rigid"
 accelerated external field (i.e. an external classical field such that
the resulting W.K.B. trajectories will be at constant Rindler position
and thus invariant under boosts) instead of the homogeneous
field which gives all trajectories with the same acceleration.
In such a case, they claim that the recoils of the accelerated system
would not destroy the coherence of the emissions since one would still have
a Rindler Killing vector guaranteeing Rindler energy conservation.
My objections to such a construction are the following:

1. The total Minkowski energy emitted is independent of the recoil.

2. If one includes the accelerating external field into the dynamics
its own recoil would not be negligible after a proper time given by
eq. (\ref{taur}) with $M$ replaced by the mass of the total system.
For these reasons
I do not consider that the "Rindler rigid" construction conceptually differs
from the one presented here.}
 lead to the
decoherence of the emissions causing these changes. This can be seen directly
from the argument of the $\delta$ function of eq. (\ref{T23})
and the spread in $\om$ ($=E^{1/2}$): when
$\om + \om^\prime >> E^{1/2}$ the overlap of the two wave packets vanishes.
This is not the case for $\langle T_{UU} \rangle_1$ since the argument
of the $\delta$ in eq. (\ref{T13}) contains $\om - \om^\prime$.

\vskip .3 truecm

We conclude this paper by three remarks.

1. We mention the work\cite{CW} of Chung and Verlinde
who attempt to take into account the recoils of a non
inertial mirror which follows,
in the mean, the trajectory $aV=-e^{-aU}$, (compare with eq. (\ref{saddtau})),
which leads to a thermal flux as in the case of a collapsing
black hole\cite{Hawk}.
The difficulty of their approach is compounded by the fact that the quantum
source of the recoils is not the individual quantum emission acts (as
in this paper, see
eq. (\ref{calB}))
but rather the quantized version of the mean energy momentum (i.e. the mean
flux expressed in terms of the quantized trajectory).

2. We point out that, in the semiclassical
treatment\cite{Bard}\cite{Massar2}\cite{PP}, the black hole loses mass
through the absorption of
a negative mean flux which crosses its future horizon.
Furthermore, this mean
negative flux can be
viewed as arising from the interferences of states with different
local particle number\cite{MPB2}. We now recall that
in the accelerating situation, upon considering the quantum recoils of
the system, the decoherence of successive emissions
washes out the interferences leading to a negative
mean flux after a logarithmicly
short time.
Since the "transplanckian" frequencies\cite{THooft}\cite{Jacobson1}\cite{EMP}
encountered in black hole evaporation
have their origin in the exponentially varying Doppler shift,
in a manner similar\cite{PaBr1} to that of the accelerated detector,
one may call into question the validity of the mean negative flux crossing the
future horizon when quantum
gravitational recoils will be
taken into account.

\newpage
3. We
remark
that the recoil of the accelerated system which follows from
strict momentum conservation bears some similarities with the recoil
of the gravitational part of the universe induced by some matter change and
 taken into account through the
quantum
character of the wave-function of gravity $+$ matter
 (solution of the Wheeler de Witt
equation in a mini superspace reduction). In particular,
the emergence of the "Banks"
time\cite{Banks}\cite{BV}\cite{BPS}\cite{Parethese}
 occurs for exactly
the same reasons that have lead to the proper time parametrization of
the
$\varphi_2$ term in eq. (\ref{phase2}), i.e. $\varphi_2(t^*)=\Delta m
\tau^*$.

\vskip 1. truecm

{\bf Acknowlegdments	} I thank S. Massar for numerous clarifying discussions
and for the precious remark that the norm of the amplitude can be
exactly computed by making no appeal to the W.K.B nor to the saddle
point approximation.
 I also wish
to thank D. Sciama for convincing encouragements to complete this work
and P. Grove and D. Raine for many useful discussions during two weeks
spent at SISSA in September 1994.
I also thank O. Fonarev for help to resolve some algebraic
difficulties. I'm grateful to SISSA for their hospitality during which
I enjoyed discussions with D. Sciama, P. Grove and D. Raine.


\end{document}